\documentclass{mn2e}
\usepackage{graphicx}

\def\go{
\mathrel{\raise.3ex\hbox{$>$}\mkern-14mu\lower0.6ex\hbox{$\sim$}}
}
\def\lo{
\mathrel{\raise.3ex\hbox{$<$}\mkern-14mu\lower0.6ex\hbox{$\sim$}}
}
\def\simeq{
\mathrel{\raise.3ex\hbox{$\sim$}\mkern-14mu\lower0.4ex\hbox{$-$}}
}

\def\etal{{et al.\ }}

\begin{document}

\title[GRB050223: A faint Gamma-Ray Burst discovered by {\it Swift}]
{GRB050223: A faint Gamma-Ray Burst discovered by {\it Swift}}
\author[K.L. Page \etal]{K.L. Page$^{1}$, E. Rol$^{1}$, A.J. Levan$^{1}$,
B. Zhang$^{2}$, J.P. Osborne$^{1}$, P.T. O'Brien$^{1}$, \and
A.P. Beardmore$^{1}$, D.N. Burrows$^{3}$, S. Campana$^{4}$,
G. Chincharini$^{4,5}$, J.R. Cummings$^{6}$,  \and G. Cusumano$^{7}$, N. Gehrels$^{6}$, P. Giommi$^{8}$, M.R. Goad$^{1}$,
O. Godet$^{1}$, V. Mangano$^{6}$,\and G. Tagliaferri$^{4}$ \& A.A. Wells$^{1}$\\
$^{1}$ X-Ray and Observational Astronomy Group, Department of Physics \&
  Astronomy, University of Leicester, LE1 7RH, UK\\
$^{2}$ University of Nevada, Box 454002, Las Vegas, NV  89154-4002, USA\\
$^{3}$ Department of Astronomy \& Astrophysics, 525 Davey Lab, Pennsylvania
  State University, University Park, PA 16802, USA\\
$^{4}$ INAF - Osservatorio Astronomico di Brera, Via Bianchi 46, 23807 Merate, Italy\\
$^{5}$ Universit{\` a} degli studi di Milano-Bicocca, Dipartimento di Fisica,
  Piazza delle Scienze 3, I-20126 Milan, Italy\\
$^{6}$ NASA Goddard Space Flight Center, Greenbelt, MD 20771, USA\\
$^{7}$ INAF - Istituto di Astrofisica Spaziale e Fisica Cosmica Sezione di
  Palermo, Via Ugo La Malfa 153, 90146 Palermo, Italy\\
$^{8}$ ASI Science Data Center, via Galileo Galilei, 00044 Frascati, Italy 
\\
}

\label{firstpage}

\maketitle

\begin{abstract}

GRB050223 was discovered by the {\it Swift} Gamma-Ray Burst Explorer on 23
February 2005 and was the first Gamma-Ray Burst to be observed by both {\it
  Swift} and {\it XMM-Newton}. At the time of writing (May 2005), it
has one of the faintest GRB afterglows ever observed.
The spacecraft could not slew
immediately to the burst, so the first X-ray and optical observations occurred
approximately 45 minutes after the trigger. Although no optical emission was
found by any instrument, both {\it Swift} and {\it XMM-Newton} detected the
fading X-ray afterglow. Combined data from both of these observatories show
the afterglow to be fading monotonically as 0.99$^{+0.15}_{-0.12}$ over a time
frame between 45 minutes to 27
hours post-burst. Spectral analysis, allowed largely  by the higher
through-put of {\it XMM-Newton}, implies a power-law with a slope of
$\Gamma$~=~1.75$^{+0.19}_{-0.18}$ and shows no
evidence for absorption above the Galactic column of 7~$\times$~10$^{20}$
cm$^{-2}$.

From the X-ray decay and spectral slopes, a low
electron power-law index of $p$~=~1.3--1.9 is derived; the slopes also imply
that a jet-break has not occured up to 27 hours after the burst.  The faintness of GRB050223 may be due to a large jet 
opening or viewing angle or a high redshift.


\end{abstract}

\begin{keywords}
gamma-rays:bursts

\end{keywords}

\date{Received / Accepted}

\section{Introduction}
\label{intro}

The {\it Swift} Gamma-Ray Burst Explorer (Gehrels \etal 2004) was launched on
20th November 2004. It is a multi-wavelength observatory, covering the gamma-ray, X-ray and
UV/optical bands. The observatory is designed to slew rapidly and autonomously to
point narrow-field instruments (the X-ray and Ultra-Violet/Optical Telescopes - XRT
and UVOT, respectively) towards any Gamma-Ray Bursts (GRBs) detected by the
Burst Alert Telescope (BAT). This allows prompt observations of the
afterglow on a timescale of minutes, much more quickly than was previously
feasible on a regular basis. The on-board instruments are described in detail
in Barthelmy (2004, 2005; BAT), Burrows \etal (2004, 2005; XRT) and Roming
\etal (2004, 2005; UVOT).

{\it Swift} is significantly more sensitive to detecting GRBs than previous
instruments capable of providing 
rapid, accurate (to within a few arcmin) localisations (e.g.,  {\it HETE-2} and
{\it BeppoSAX}). Thanks to its rapid repointing capability, {\it Swift} is
also able to observe afterglows at early times. Since GRB afterglows fade
rapidly, this ensures they are observed at their brightest, allowing {\it
  Swift} to detect fainter afterglows and
thus look further down the
GRB afterglow luminosity function than has previously been
possible. Investigating the faint end of this function is of particular
importance in understanding the structure of the bursts themselves.
Faint bursts may be manifestations of many different effects, such as a
large luminosity distance [{\it Swift} should be able to detect
bursts out to z~$\sim$~15--20
(Lamb \& Reichart 2000)] or differences in the fireball emission (shock
generation, jet structure). Alternatively, they could be due to a separate
population of low luminosity, relatively nearby (z~$<$~0.2) bursts
(e.g. Sazonov et al. 2004). The combined study of the prompt and afterglow emission of these
bursts will make it possible to distinguish between these possibilities. 

Here results of {\it Swift} and {\it XMM-Newton} observations of GRB050223,
which has one of the faintest
X-ray afterglows to date, are presented and constraints are placed on some of the
burst and afterglow parameters.

\section{Observations}
\label{obs}

GRB050223 (Swift Trigger 106709) was detected by the {\it Swift} BAT at
03:09:06 UT on 23 February 2005 (Mitani \etal 2005), at a location of
RA(J2000)~=~18$^{h}$05$^{m}$34$^{s}$, Dec(J2000)~=~$-$62$^{\circ}$28$^{'}$52$^{''}$, with an uncertainty of
4~arcmin; the burst was also detected by  {\it INTEGRAL} (Mereghetti \etal
2005). Because of the Earth-limb
constraint, the {\it Swift} spacecraft could not slew to the BAT position until 03:44 UT,
at which point the observatory was in the South Atlantic Anomaly (SAA). The
XRT began collecting data upon exiting the SAA, at
03:56:37 UT. An uncatalogued X-ray source was identified at RA(J2000)~=~18$^{h}$05$^{m}$32.6$^{s}$,
Dec(J2000)~=~$-$62$^{\circ}$28$^{'}$19.7$^{''}$, with an uncertainty of 8~arcsec (Giommi
\etal 2005); this is 33~arcsec from the BAT position. The UVOT began observations slightly before the XRT, at 03:55:28 UT. 

Since this GRB was detected during the calibration phase of {\it Swift}, the
XRT was in Manual State, where data-mode switching is not automatically enabled; there were, therefore, no automatic alerts sent out
via TDRSS (the Tracking and Data-Relay Satellite System). Also, during the initial
observation all data were obtained in Photon Counting (PC) mode, rather than
the standard cycle starting with an Image Mode frame.

{\it Swift} software version 1.2 was used to process the XRT and BAT data. 
The BAT files were processed using the latest version (2.17) of the analysis
script, which produces mask-weighted spectra and light-curves. For the XRT,
events below a threshold of 80 DN (approximately 0.2~keV) were filtered out
and the bad pixels removed. This method ensures that the event-file is as
clean as possible, removing the effects of the sunlit Earth, and is the default pipeline method for later releases of the software.


Source and background spectra
were then extracted using a circular
region of radius 15~pixels (1~pixel~=~2.36~arcsec). 
Only grade 0 events were used for the XRT PC mode spectra, since
the response matrix (RMF) for these single pixel events (swxpc0$\_$20010101v006.rmf)
was the best calibrated at the time of analysis; using all calibrated grades (0--12) did not
significantly improve the statistics. Grades 0--12 were used for
the light-curves, however. The {\sc ftool} {\bf
  xrtmkarf} was used to generate suitable ancillary response function (ARF)
files for the spectral fitting.

{\it XMM-Newton} also
observed the field of GRB050223 (Gonzalez-Riestra \etal 2005; Rodriguez 2005; De Luca \&
Campana 2005).
SAS v6.1 was used for these data, choosing patterns (equivalent to {\it Swift}
grades) 0--12 for
MOS and 0--4 for PN. Background light-curves showed 
frequent flaring for the later {\it XMM-Newton} observation, particularly in
the PN data, so a small source extraction radius
(35~arcsec) was used in addition to screening out the worst of the
background contribution. The SAS tasks {\bf rmfgen} and {\bf arfgen} were then run
to produce the RMF and ARF files respectively.

All spectra were grouped to a minimum of 20 counts per bin, in order to
facilitate $\chi^2$ fitting in {\sc xspec} v11.3.1. Throughout this
Letter, errors are given at the 90~per~cent level (e.g.,
$\Delta\chi^{2}$~=~2.7 for one degree of freedom).

\subsection{Gamma-ray data}

\begin{figure}
\begin{center}
\includegraphics[clip, width=6cm,angle=-90]{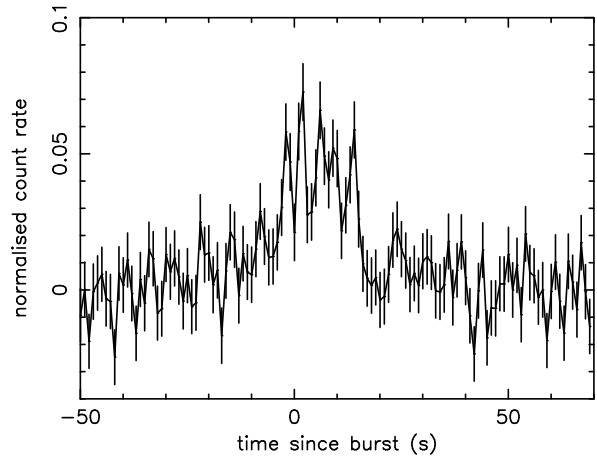}
\caption{The BAT light-curve, over 15-350~keV, with 1-second binning.}
\label{batlc}
\end{center}
\end{figure}

The BAT light-curve of GRB050223 shows a slow rise and fall, superimposed by
several short peaks (Mitani \etal 2005; Figure~\ref{batlc}). T$_{90}$ for this burst is 
23~seconds, while the peak flux, over a 1-second interval, was 0.8
photon~cm$^{-2}$~s$^{-1}$ (15--350~keV; Mitani \etal 2005). 

The {\it INTEGRAL} IBIS/ISGRI (Imager on-Board the {\it INTEGRAL} Satellite/{\it INTEGRAL} Soft
Gamma-Ray Imager) instrument also detected GRB050223, measuring a
peak flux (1-second integration) of 0.6 photon~cm$^{-2}$~s$^{-1}$ over
20-200~keV (Mereghetti \etal 2005).

\begin{figure}
\begin{center}
\includegraphics[clip, width=6cm,angle=-90]{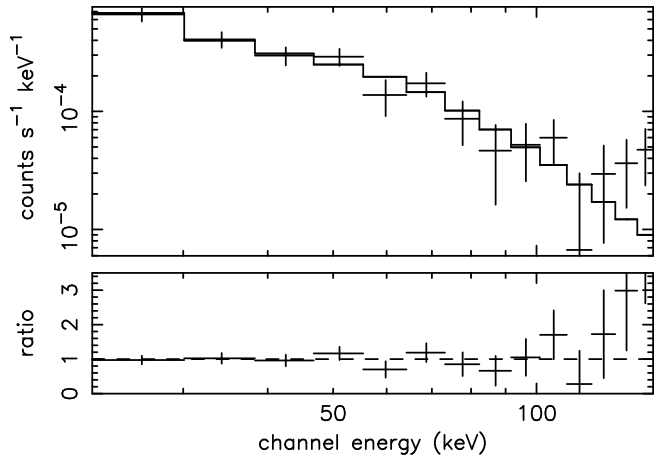}
\caption{The BAT spectrum can be fitted with a simple power-law model, with $\Gamma_\gamma$~$\approx$~1.85.}
\label{batspec}
\end{center}
\end{figure}

A single power-law gave a good
fit ($\chi^2$/dof~=~48/57; Figure~\ref{batspec}) for $\Gamma_\gamma$~=~1.85~$\pm$~0.19\footnote{When
  considering spectral slopes in X-ray astronomy, the convention is to give
  the value as $\Gamma$, the photon index, where f(E)~$\propto$~E$^{-\Gamma}$;
  f(E) in units of photon~cm$^{-2}$~s$^{-1}$.}, which was not
improved upon by using the Band model (Band \etal 1993). The energy
fluence over 15--350~keV was 9.69~$\times$~10$^{-7}$ erg~cm$^{-2}$, placing it
in the lowest third of the {\it Swift}-measured fluence distribution. 

\subsection{X-ray Data}

Table~\ref{exp} lists the times and durations of the X-ray data
obtained from {\it Swift} and {\it XMM-Newton}. 
All the useful {\it Swift} data were obtained in PC mode, both for the
initial (three orbits of data) and second (seven orbits when settled
on the source) observations. The {\it XMM-Newton} MOS1 and
MOS2 data were checked for consistency and then co-added for subsequent
analysis. The {\it XMM-Newton} PN data are more badly affected by the high background, so
are not presented here, but the results are in agreement with the MOS.

\begin{table*}
\begin{center}
\caption{Exposure times for the {\it Swift}-XRT (PC mode) and {\it
    XMM-Newton}-MOS data.  Burst trigger time from the BAT was
    2005-02-23T03:09:06 UT. The first {\it Swift} observation corresponds to
    sequence number 00106709000; the second - 00106709001. The first and
    second {\it
    XMM-Newton} observations correspond to before and after a ground station
    outage between 2005-02-23T16:54 UT and  2005-02-23T18:56 UT.} 
\label{exp}
\begin{tabular}{p{3.0truecm}p{1.5truecm}p{1.5truecm}p{2.7truecm}p{2.7truecm}}
\hline
Instrument & Observation & Orbit & Start time  &  End time \\
& & & (s after BAT trigger) & (s after BAT trigger)  \\
\hline
{\it Swift} XRT & 1 & 1 & 2847 & 3973\\
{\it Swift} XRT & 1 & 2 & 9150 &  9710\\
{\it Swift} XRT & 1 & 3 & 14665 & 15475\\
{\it Swift} XRT & 2 & 1-7 & 38265 & 73530 \\
{\it XMM} MOS1/MOS2& 3 & - & 35746/35745 & 49526/49533  \\
{\it XMM} MOS1/MOS2& 4 & - & 57450/57527 & 96452/96456 \\
\hline
\end{tabular}
\end{center}
\end{table*}
\subsubsection{Light-curve Analysis}

Because of the location of GRB050223, most of the {\it Swift}-XRT pointings were close to
the Earth limb (small `Bright Earth' angles). This led to a high optical
background in the field of view which, together with the afterglow being faint,
complicated the X-ray data analysis. 

\begin{figure}
\begin{center}
\includegraphics[clip, width=6.2cm,angle=-90]{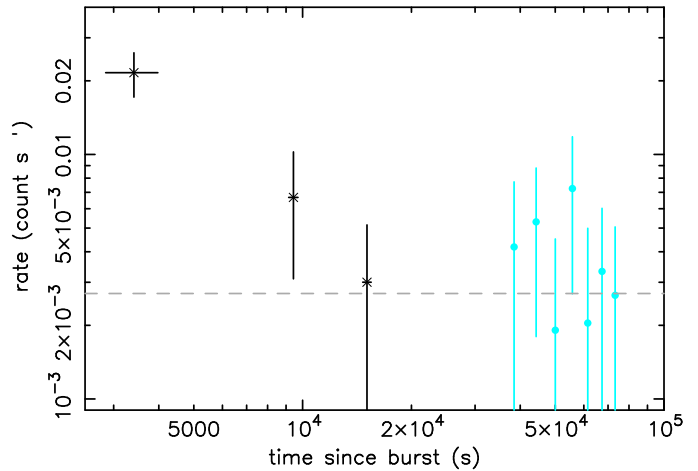}
\caption{{\it Swift}-XRT light-curves for each orbit of GRB050223 data; the first three points (marked with X-symbols) show data
 from sequence number 00106709000; the following seven points (circles) are
 from 00106709001. The dashed horizontal line shows the background level; with
 the exception of the data from the first two orbits, the burst is not
 significantly detected above the background in individual orbit segments.}
\label{decay}
\end{center}
\end{figure}

Light-curves were extracted for each individual orbit of data. Because of the
faintness of the afterglow, there were very few counts in each of the orbit
bins (Table~\ref{counts}) so, in order to improve the statistics, a large
background region was used (circle of radius 60~pixels) and the number of background counts scaled down to the size of
the source region (radius 15~pixels). The count-rates were corrected for the
fractional exposure where required. As Figure~\ref{decay} shows, only the first two orbits of
data show count rates significantly above the background level of
around 2.7~$\times$10$^{-3}$ count~s$^{-1}$ (within the
15-pixel radius circle). Considering the second
{\it Swift} observation as a whole, the source is detected at the 3$\sigma$
level (using the {\bf detect}
command in {\sc ximage}).

\begin{table}
\begin{center}
\caption{The number of source counts (to one decimal place) for each {\it Swift}-XRT observation,
  integrated over each orbital time bin. The last column gives the fraction of
  the time-bin during which data were actually collected.} 
\label{counts}
\begin{tabular}{p{1.5truecm}p{1.0truecm}p{1.0truecm}p{1.5truecm}p{1.5truecm}}
\hline
Observation number & Orbit number & Source counts & Time-bin (s) & Exposure fraction\\
\hline
1 & 1 & 24.3 & 1125 & 0.89\\
1 & 2 & 3.7 & 560 & 0.96\\
1 & 3 & 2.4 & 810 & 0.84\\
2 & 1 & 1.7 & 400 & 1.0\\
2 & 2 & 2.6 & 495 & 1.0\\
2 & 3 & 0.7 & 385 & 1.0\\
2 & 4 & 2.8 & 380 & 1.0\\
2 & 5 & 0.7 & 340& 1.0\\
2 & 6 & 1.7 & 525& 1.0\\
2 & 7 & 1.6 & 590 & 1.0\\
\hline
\end{tabular}
\end{center}
\end{table}

\begin{figure}
\begin{center}
\includegraphics[clip, width=6cm,angle=-90]{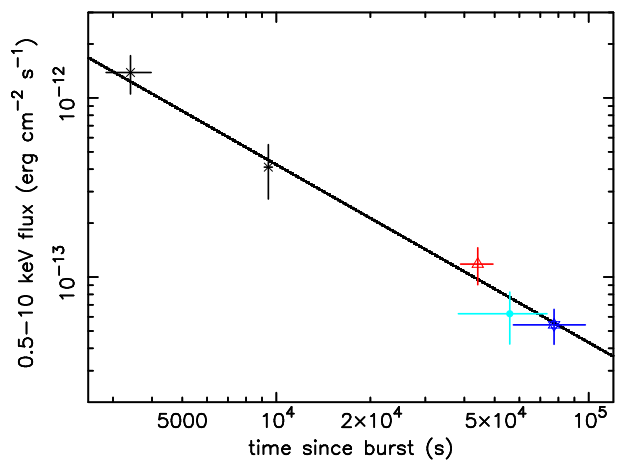}
\caption{Plotting the light-curve in terms of flux, the {\it XMM-Newton}
  measurements can be compared with those from {\it Swift}. The X-symbols show
  the first two orbits of {\it Swift} data (within observation one) and the circle the whole of the second
  {\it Swift} observation, while the {\it
  XMM-Newton} points are marked by triangles and stars (observations 3 and 4
  respectively). The model shown is a decay slope of 0.99.}
\label{fluxdecay}
\end{center}
\end{figure}

In order to compare the data from {\it XMM-Newton} (Obs. ID 0164570601) with the {\it Swift}
results, the light-curve has to be plotted in terms of flux using the spectral
fit given in Section~\ref{spec}, rather than
count-rate, because of the differences between the two instruments. The
background for the {\it XMM-Newton}-MOS detectors was checked and found
to be about a third the count-rate of the source before the ground station
outage, and about half afterwards, so the burst is clearly
detected. A combined light-curve of the {\it Swift} and {\it XMM-Newton}
observations is plotted in Figure~\ref{fluxdecay}, showing a decay
slope\footnote{f(t,$\nu$)~$\propto$~t$^{-\alpha}\nu^{-\beta}$ where $\beta$~=~$\Gamma-$1.} of
$\alpha$~=~0.99$^{+0.15}_{-0.12}$. The second {\it Swift} observation occured
simultaneously with the {\it XMM-Newton} observations, with the values from the
different satellites being in good agreement.


\subsubsection{Spectral Analysis}
\label{spec}

\begin{figure}
\begin{center}
\includegraphics[clip, width=6cm,angle=-90]{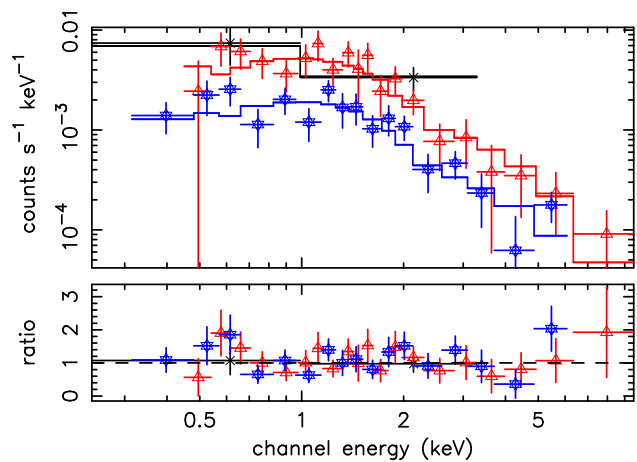}
\caption{A power-law fit ($\Gamma$~$\approx$~1.75) to the joint {\it Swift}-XRT (X-symbols) and {\it
    XMM-Newton} (triangles and stars for observations three and four respectively) data.  The {\it Swift} spectrum was
    formed from the first three orbits of data. The spectrum from the second
    {\it Swift} observation consists of a single bin of data, so has not been included.
   }
\label{spectra}
\end{center}
\end{figure}

Because of the faintness of the X-ray afterglow and the high optical
background, the {\it Swift}-XRT spectrum is of low statistical quality. However,
simultaneously fitting
the spectrum, derived from the three orbits in the first observation, and the co-added {\it
  XMM-Newton}-MOS spectra produces a good fit ($\chi^2$/dof~=~28/34; Figure~\ref{spectra}) for a
single power-law of $\Gamma$~=~1.75$^{+0.19}_{-0.18}$
 absorbed by the Galactic
column of 7~$\times$~10$^{20}$~cm$^{-2}$ (Dickey \& Lockman 1990), with a
different constant of normalisation
between the individual spectra. No change in spectral shape is found between
the {\it XMM-Newton} spectra.

Note that the spectrum is shown in detected counts~s$^{-1}$~keV$^{-1}$ for
each of the instruments. Thus, while the {\it XMM-Newton} spectrum may have a
higher count-rate, due to the higher throughput, this does not correspond to
an increased flux.

The unabsorbed fluxes (0.5--10~keV) for
observations one, three and four (as named in Table~\ref{exp}) were found to be (8.18$^{+3.32}_{-2.74}$)~$\times$~10$^{-13}$,
(1.18$^{+0.19}_{-0.36}$)~$\times$~10$^{-13}$ and (5.42$^{+0.98}_{-1.44}$)~$\times$~10$^{-14}$ erg~cm$^{-2}$~s$^{-1}$ respectively.

\subsection{UV and Optical data}

Neither the {\it Swift}-UVOT (Gronwall \etal 2005) nor the {\it XMM-Newton}
Optical Monitor (Blustin \etal 2005) detected a source at the position of the
X-ray afterglow. As mentioned above, the UVOT observation started about 46
minutes after the BAT trigger, due to the delayed slew; the {\it
  XMM-Newton}-OM data were collected 11~hours after the trigger.

No new sources were identified by {\it ROTSE-III} (to a limiting unfiltered
magnitude of 18 from
approximately a minute after the burst; Smith 2005), the Mount John
University Observatory (to R~=~20.5, 10~hours after the burst; Gorosabel \etal 2005) or the {\it PROMPT} robotic telescope
array (limiting magnitude of $\approx$~21 for Rc, V and Ic filters, with the
mean time for these observations being 4--5~hours after the trigger;
Nysewander \etal 2005). 


\section{Discussion}
\label{disc}

GRB050223 has, at the time of writing (May 2005), one of the faintest GRB
X-ray afterglows observed by {\it Swift}; comparison with figure 1 of Piro
(2004) shows the 11~hour flux of GRB050223 to be below all those detected by
{\it BeppoSAX}.

\subsection{Afterglow models}

Three GRB afterglow models are initially considered, as summarised by
Zhang \& M{\'e}sz{\'a}ros (2003). The `ISM' model has a
fireball expanding into the (homogeneous) interstellar medium (Sari,
Piran \& Narayan 1998), while, in the `Wind' model, the fireball expands into
a wind environment, with the density, $\rho~\propto$~r$^{-2}$ (Chevalier
\& Li 1999). In these models the beaming angle (1/$\Gamma_{0}$, where
$\Gamma_{0}$ is the Lorentz factor; this is simply the cone into which the
emission is beamed due to relativistic effects) is less than any jet opening 
angle. As the jet slows down, 1/$\Gamma_{0}$ will become larger; when it
becomes equal to the opening angle, a transition, known as the jet-break, is
seen. At this point, the emission observed decreases due to both the edge
effect (less emission per unit solid angle is seen) and the sideways spreading of the
causally-connected region. These effects may not happen simultaneously, but
are thought to be close in time (Panaitescu \&  M{\'e}sz{\'a}ros 1999; Sari
Piran \& Halpern 1999). The third model is for post jet-break evolution, when the finite
angular extent of the jet dominates (Sari \etal 1999), which is
valid for both ISM and wind cases. 

The reasonable assumptions that the X-ray afterglow lies above the
synchrotron injection frequency ($\nu_m$) and that during the XRT 
observations, hours after the GRB, slow cooling is effective 
(i.e.~$\nu_{X} > \nu_c$, the X-ray frequency is greater than the cooling 
frequency) are made. Then the afterglow temporal decay and spectral indices 
($\alpha$~=~0.99$^{+0.15}_{-0.12}$ and $\Gamma$~=~1.75$^{+0.19}_{-0.18}$) 
indicate an electron power-law index $p$~=~1.3--1.9 for a spherical blast-wave 
since $\alpha$~=~(3$p$+10)/16 for the ISM case, ($p$+6)/8 for wind cooling 
and ($p$+6)/4 for the jet-dominated case, while $\Gamma$ is given by 1+($p$/2) 
for each (Dai \& Cheng 2001).

The data are consistent with either an ISM or wind regime.
For jet-dominated evolution the high frequency emission falls off as 
t$^{-(p+6)/4}$, much steeper than
the decay observed in GRB050223. As might be expected for these relatively early 
observations, our spectral and temporal slopes are inconsistent with
post jet-break evolution.

A value of $p$ less than two is not generally thought to be physical (e.g.,
Panaitescu \& Kumar 2001), although possible ways to generate such a flat
spectrum have been suggested (e.g., Bykov \& M{\'e}sz{\'a}ros 1996).
A similarly low value for $p$ was among the possibilities for GRB050128 
(Campana \etal 2005) if the observed change in slope of the decay-curve was caused by 
a jet-break in that burst.

A jet-break in the light-curve for a large opening angle would naturally occur 
at a late time (Piran 1999). A late jet-break
is in agreement with the analysis above, which indicates that the
outflow prior to the jet-break is being observed, with no indication
of such a break up to at least 10$^{5}$ seconds. Jet-breaks are frequently
observed at longer than a day after the burst (see, e.g., Frail \etal 2001),
so this is not unusual. A large opening angle could 
also explain the relative faintness of the X-ray afterglow and the BAT fluence 
being at the lower end of the {\it Swift} fluence distribution.

The GRB jet opening angle can be estimated to be 
$\theta_j$~$\approx$~0.35--0.4~rad using the observed correlations of 
gamma-ray fluence and X-ray afterglow decay index with a jet opening angle 
measured by jet-break times for ten GRBs by Liang (2004). 
Our jet angle estimate is relatively large
compared to the sample of Frail \etal (2001). 
It should, however, be noted that
the Liang relationships were derived from a sample of only ten bursts
and doubts about their general applicability remain. 
Also, Bloom, Frail
\& Kulkarni (2003) list bursts (e.g., GRB000418 and GRB021004) which are
bright, yet have larger than typical opening angles.

If GRB050223 produced a structured jet [that is,
$\Gamma$($\theta$)~$\propto$~$\theta^{-q}$], then the faintness seen here could be 
due to a large viewing angle from the jet axis. In this case the viewing angle 
corresponds to a low energy density in the jet. The absence of a jet-break 
before one day in our data is consistent with an off-axis viewing angle
(Zhang \& M{\'e}sz{\'a}ros 2002; Rossi, Lazzati \& Rees 2002).

Alternatively, the observed low afterglow flux and prompt fluence could be
explained by GRB050223 being at high redshift. In this case, any jet-break is
delayed by a factor proportional to $1+z$. Indeed, {\it Swift} bursts
to date are on average fainter than those detected by {\it BeppoSax} and {\it
 HETE-2} (Piro 2004; Berger \etal 2005) and the median redshift of the six {\it Swift} bursts for 
which it has been measured so far is large, at z~=~2.4, compared to a median 
z~=~1.0 for non-{\it Swift} bursts\footnote{Values taken from Jochen 
Greiner's website at http://www.mpe.mpg.de/$\sim$jcg/grbgen.html}. 

\section{Summary}
\label{sum}

Observations by {\it Swift} and {\it XMM-Newton} have shown 
GRB050223 to have faint prompt gamma-ray and X-ray afterglow emission.
The X-ray data agree with the standard stellar wind and constant circumstellar
density afterglow models if the electron power-law index, $p$~=~1.3--1.9. 
A jet-break does not appear to have occured up to one day after the burst.
The faintness of GRB050223 may be due to a large jet opening or viewing angle,
or a high redshift.

\section{ACKNOWLEDGMENTS}
\label{ack}

We thank Don Lamb for useful comments on the paper and Jochen Greiner for his
GRB afterglows website.
This work is supported at the University of Leicester by the Particle Physics
and Astronomy Research Council (PPARC), at Penn State by NASA contract
NAS5-00136 and in Italy by funding from ASI (contract number I/R/039/04). JRC is supported by a National Research Council
Associateship award at NASA's Goddard Space Flight Center.

\end{document}